\documentclass{amsart}
\usepackage{a4wide}
\usepackage{amssymb}
\newtheorem{definition}{Definition}
\newtheorem{theorem}{Theorem}
\newtheorem{lemma}[theorem]{Lemma}
\def\RR{\mathbb{R}}
\begin{document}

\begin{abstract}{ We consider Burgers equation forced by a brownian in
    space and white noise in time process $\partial_t u +
    \frac{1}{2} \partial_x(u)^2 = f(x,t)$, with  $E( f(x,t) f(y,s) ) =
 \frac{1}{2}(|x|+|y|-|x-y|)\delta(t-s)$ and
we show that there are Levy processes solutions, for which we give the
evolution equation of the characteristic exponent. In particular we give the
explicit solution in the case $u_0(x)=0$.} 
\end{abstract}

\title{Levy solutions of a randomly forced Burgers equation}
\author[M-L. Chabanol]{Marie-Line Chabanol}
\address{Institut de Math\'ematiques de Bordeaux\\UMR 5251 CNRS-Universit\'e Bordeaux1\\France}
\email{Marie-Line.Chabanol@math.u-bordeaux1.fr}
\author[J. Duchon]{Jean Duchon}
\address{Institut Fourier (Grenoble)\\UMR 5582 CNRS-Universit\'e Joseph
Fourier\\100, rue des Math\'ematiques, B.P. 74\\
38041 Saint-Martin d'H\`eres Cedex\\France}
\email{Jean.Duchon@ujf-grenoble.fr}
\thanks{Marie-Line.Chabanol@math.u-bordeaux1.fr, Jean.Duchon@ujf-grenoble.fr}
\maketitle
\section{Introduction}
We consider the randomly forced Burgers equation $\partial_t u + \frac{1}{2} 
\partial_x (u)^2 = f(x,t)$, where $f$ satisfies $E( f(x,t) f(y,s) ) =
 A(x,y)\delta(t-s)$ with $A(x,y) = \frac{1}{2}(|x|+|y|-|x-y|)$ : $f$ 
 is a brownian in space/ white noise in time
process (the brownian is considered on the whole space, and is obtained with
two independent brownians, one on the left and one on the right). 
Burgers equation has originally been introduced as a $1D$ model of
turbulence. It is certainly too crude, but it is still a good idea
when trying to find new approaches to the ``real'' 3D turbulence as
given by Euler equation, to start by looking at Burgers. In this point
of view, the randomly forced Burgers equation is a crude
simplification on the way to the description of forced turbulence.
A variant of our problem was considered by \cite{EvdE}, where they considered
a gaussian in space/white noise in time forcing, satisfying
$E(f(x,t) f(y,s)) = A(x-y) \delta(t-s)$ and obtained all the
hierarchy of $n$-point densities evolution equations.
We will proceed quite differently, basing our study on the
 result  shown by Carraro and Duchon in \cite{CD}, and
by Bertoin in \cite{Ber},
 that Levy processes with negative jumps are 
statistical solutions (in a sense that we will recall later) of the Burgers
equation without any forcing. This is actually a particular case of
the fact that the Burgers equation conserves the Markovian (in space)
property of processes (\cite{ChD}). We will show here  in the same spirit
%see using basically the same method as in \ref{CD}  
that 
the brownian in space/white noise
in time forcing allows to keep the Levy property. Moreover, our method allows 
us to write explicitly the exponent of the Levy process that one gets
when starting from $u_0(x)=0$.   

\section{Statistical solutions}

We will closely follow~\cite{CD}  (see also~\cite{RRinFriedSerre}).
Let $E$ be the space of c\`adl\`ag real functions. We will call 
$\mathcal C(E)$ the smallest $\sigma$-algebra such that for each $x \in  \RR$, $u \mapsto u(x)$ 
is measurable, and   $\mathcal C'(E)$ the smallest $\sigma$-algebra
such that for each $(x,y) \in \RR\times \RR$, $u \mapsto u(x)-u(y)$ 
is measurable.
Let $\mathcal D_0$ be the set of real $C^\infty$ functions $v$ with compact
support such that $\int_{\RR} v(x)dx = 0$.
A probability $\mu$ on $(E,\mathcal C(E))$ 
is then characterized by its characteristic
function 
\[
 v \in {\mathcal D} \mapsto \hat{\mu}(v):=\int_E \exp \, [i \int_{\RR} u(x) v(x)
dx]\, d\mu(u)
\]
whereas a probability $\mu$ on $(E,\mathcal C'E))$ 
is  characterized by its characteristic
function $ v \in {\mathcal D}_0 \mapsto \hat{\mu}(v)$.

Let $u_0$~:  $(\Omega, {\mathcal A}, P) \to E$ be a random process, defined
on some probability space, and let $\mu_0 : \mathcal C(E)\to [0,1]$  denote its
probability law : $u_0$ will be our initial condition. 
We will let $u$ evolve according to the non forced Burgers equation,
obtaining thus a family of processes $u_t$ where $u_t(x)=u(x,t)$ is
the solution of Burgers at time $t$.  Then
it is not difficult to check that the characteristic function $\hat{\mu}_t$ of 
$u_t$ verifies :
\begin{eqnarray}
\partial_t \hat{\mu}_t(v) &=&\partial_t( \int_E \exp \, [ i \int_{\RR} u(x)
v(x)dx]\, d\mu_t(u) \nonumber\\
&=& \int_E \partial_t \{\exp \, [i \int_{\RR} u(x,t) v(x)dx]\}\,
d\mu_0(u_0) \nonumber\\
&=&  \int_E  \exp \, [i \int_{\RR} u(x,t) v(x) dx)] \,
(\partial_t[i\int_{\RR}  u(x,t) v(x) dx ]) \,d\mu_0(u_0) \nonumber\\
&=& \int_E  \exp \, [i \int_{\RR}u(x) v(x)dx]\,[i \int_{\RR}
\frac{1}{2} u(x)^2 v'(x) dx] d\mu_t(u)\label{stats}
\end{eqnarray}

This motivated the definition given in \cite{CD} of a statistical solution 
of Burgers equation, as a family of random processes $u_t$ whose characteristic functions are solutions of (\ref{stats}).

Here we will
take for $u_0$ a Levy process with negative jumps, that is a
c\`adl\`ag process with stationary homogenous increments and negative
jumps, 
and we are really interested only in the law of the increments. Hence $u_0$ really defines 
a probability law on $(E,\mathcal C'E))$. But  $u\mapsto \int_{\RR} u(x)^2 v'(x)dx$ is not
$C'(E)$ measurable, hence this evolving equation for $\hat{\mu}$ makes no sense. One way around this is to take a different approach and to 
work with the Hopf-Cole construction as in
\cite{Ber}. Another way was explored in \cite{CD} :
it is based on the notion of {\em intrinsic statistical solution},
and involves working with $u(x,t) - \frac{1}{b-a}\int_a^b u(x,t)dx $ when $b-a$
gets large :

\begin{definition}
An intrinsic statistical solution of Burgers equation is a set $(\mu_t)_{t\geq
  0}$ of probabilities on $(E,\mathcal C'(E))$ such that for any $v \in {\mathcal D_0}$,  
\begin{equation*} \partial_t \hat{\mu}_t(v) =   \frac{i}{2}\lim_{a\rightarrow
    -\infty, b \rightarrow +\infty} \int_E  \left(\int_{\RR} 
  \left(u(x) -\frac{1}{b-a} \int_a^b u(y)dy\right) ^2 v'(x) dx\right)  \exp \, [i \int_{\RR}
u\, v\,] \,d\mu_t(u))  \end{equation*} 
%\label{mut}
\end{definition}

In order to work with Levy processes we will need their
characterization by means of their exponent.

\begin{definition}
If $u$ is a Levy process, its characteristic exponent  $\psi$ is defined by 
  $\forall \, x<y, \forall \, \lambda \in
\RR$~,
\begin{equation}
E\{ \exp \,i [\lambda(u(y) - u(x))] \} = \exp[(y-x)\psi(\lambda)]\label{expo}
\end{equation}

If the process is of finite variance with negative jumps, one can
use analytic continuation to define its Laplace exponent
$\phi(\lambda) :=\psi(-i\lambda)$ as a function on $\RR^+$.
\end{definition}

 In \cite{CD} it is shown that if $u_0$ is a Levy process of finite
 variance such that $\phi'_0(0) \geq 0$, and if we let $u$
evolve according to the non forced Burgers equation, $u_t$ is still a
Levy process; its 
 characteristic exponent $\psi_t$
 satisfies the
evolution equation $\partial_t \psi = i \partial_\lambda (\psi^2)$. 

 Let now
$f$ be the random forcing, 
independent on $u_0$.  We will denote by $\mu_F$ its
probability law. Restricted to events involving a fixed time, $\mu_F$ is 
the law of a brownian motion defined on $\RR$ by gluing one brownian on
$\RR^+$ and the reflection of another independent brownian on $\RR^-$.
Such a brownian on $\RR$ is a continuous process with independent homogenous
increments null at 0. 

The basic idea now is the following : if $u_t$ is a Levy process with
probability on $C'(E)$ $\mu_t$,
% and associated measure on $C(E)$ $\tilde{\mu}_t$, 
and if we let $u_t$ evolve according to our forced
Burgers equation, $u(x,t+dt)$ will be the sum of $u_{NF}(x,t+dt)$ and 
 $\sqrt{dt} B(x)$,  where $u_{NF}$ is the
process we would have obtained without any forcing, and
  $B(x)$ is a brownian motion independent of $u_{NF}$.
%Hence we will formally
%have $\tilde{\mu}_{t+dt} = \tilde{I}_{dt}({\mu}_t) * W$ where $W$ is the Wiener
%measure rescaled by an independant gaussian variable. 
%Now the key
%point is that if $\tilde{\mu}_t$ is invariant by all translations
%$(c,h)$, so is $\tilde{I}_{dt}({\mu}_t)$, as was already mentioned,
%and so will be this $\tilde{\mu}_{t+dt}$. Indeed, if $h\in \RR$, 
%the process obtained
%from $B$ by putting the new origin at $(h,B(h))$, $B_h(x) = B(x+h) -
%B(h)$ is a brownian on $\RR$. Hence using the invariance of
%$\tilde{\mu}$, if $A \in C(E)$,
%$\tilde{\mu}_{t+dt}(\{u_{NF}(.+h,t+dt)+\sqrt dt N B(.+h)+c \in A\})=
%\tilde{\mu}_{t+dt}(\{u_{NF}(.+h,t+dt) + \sqrt dt N B_h(.)
%+\sqrt t N B(h) + \sqrt dt N c \in A\})= \tilde{\mu}_{t+dt}(\{u_{NF}(.,t+dt) +
%\sqrt dt N B_h(.) \in A\}) = \tilde{\mu}_{t+dt} (\{u_{NF}(.)+\sqrt dt N
%B(.)\in A\})$. Thus, $\tilde{\mu}_{t+dt}$ should correspond to a
%probability on $C'(E)$,
 $u(x,t+dt)$ should thus still be a  process with homogenous independent
increments, that is, a Levy process with exponent the sum of the exponents.
 Since the characteristic exponent of a
brownian is $\frac{x^2}{2}$, and since the evolution equation for
the characteristic exponent in case of no forcing is $\partial_t \psi(t,w) =
i \partial_w \psi(t,w) \psi(t,w)$, the evolution equation of the
characteristic exponent of $\mu$ with random forcing should be 
$\partial_t \psi(t,w) =
i \partial_w(\psi^2) + \frac{w^2}{2}$.  
 
One can also get this result in a more rigorous way by 
working with characteristic functions. Let us first define 
statistical solutions for the forced Burgers equation. 
Let $u(x,t)$ be a (weak) solution of our forced Burgers equation
with $u(.,0)=u_0$, $u(.,t) \in E$ for $t>0$. 
%and everything makes sense in the following calculation ~: 
%integrability, and differentiability with respect to $t$. 
Let $\mu_t$ denote the law of $u(.,t)$ on $\mathcal C(E)$.
% and let $\tilde{\mu_t}$
%be the measure on $\mathcal C(E)$ defined as before.
%Suppose first that $\tilde{\mu}_t$ is actually a probability on $E$.
Using Ito's formula, one then gets for each  $v$ in ${\mathcal D}$ : 
\begin{eqnarray*}\label{stats2}
\partial_t \hat{{\mu}}_t(v) &=& \!\!\int_E \partial_t \{\exp \, [ i \int_{\RR} u(x)
v(x)dx]\}\, d{\mu}_t(u)\\
&=&\!\! \int_F\!\!\int_E \partial_t \{\exp \, [i \int_{\RR} u(x,t) v(x)dx]\}\,
d{\mu}_0(u_0) d\mu_F\\
&=& \!\!\int_F \!\!\int_E  \exp \, [i \int_{\RR} u(x,t) v(x) dx)] \,
\left(\partial_t[i\int_{\RR}  u(x,t) v(x) dx ]\right.\\
& &
 \left.-\frac{1}{2}\int\!\!\int_{\RR\times \RR} v(x)v(y) \langle f(x,t) f(y,t)\rangle_tdxdy\right) 
\,d{\mu}_0(u_0) d\mu_F\\
%&=& \int_F\int_E  \exp \, [i \int_{\RR}u(x,t) v(x)dx]\,(i \int_{\RR}
%\frac{1}{2} u(x,t)^2 v'(x) dx \\ 
%&& + i \int_{\RR} f(x,t) v(x)dx - \int_{\RR \times \RR} v(x)v(y)
%\langle f(x,t) f(y,t)\rangle_t
%dxdy) \,d{\mu}_0(u_0) d\mu_F %\mbox{ where we used Ito's formula }
%\\
&=& \!\! \int_E  \left(\int_{\RR} i\frac{1}{2} u(x)^2 v'(x) dx - 
\frac{1}{2}\int\!\!\int_{\RR\times \RR} v(x)v(y) A(x,y)dxdy\right)\, \exp \, [i \int_{\RR}
u\, v\,] \,d{\mu}_t(u)
\end{eqnarray*}

Hence our definitions :

\begin{definition}
A statistical  solution of the forced Burgers equation (*) 
is a set $(\mu_t)_{t\geq
  0}$ of probabilities on $(E,\mathcal C(E))$ such that for any $v \in {\mathcal D}$,  
\begin{equation*} \partial_t \hat{\mu}_t(v) =   \int_E  \left(\int_{\RR} i\frac{1}{2}
  u(x) ^2 v'(x) dx\right)  \exp \, [i \int_{\RR}
u\, v\,] \,d\mu_t(u)- \left(\frac{1}{2}\!\int \!\!\int_{\RR\times
    \RR} v(x)v(y) A(x,y)dxdy \right) \hat{\mu}_t(v) \end{equation*}
\end{definition}

\begin{definition}

An intrinsic statistical solution of the forced Burgers equation (*) 
is a set $(\mu_t)_{t\geq
  0}$ of probabilities on $(E,\mathcal C'(E))$ such that for any $v \in {\mathcal D_0}$,  
\begin{eqnarray} \partial_t \hat{\mu}_t(v) &=&   \lim_{a\rightarrow
    -\infty, b \rightarrow +\infty} \int_E  \left(\int_{\RR} \frac{i}{2}
  \left(u(x) -\frac{1}{b-a} \int_a^b u(y)dy\right) ^2 v'(x) dx\right)  \exp \, [i \int_{\RR}
u\, v\,] \,d\mu_t(u)\nonumber  \\
&&- \left(\frac{1}{2}\int \!\!\int_{\RR\times
    \RR} v(x)v(y) A(x,y)dxdy \right)\hat{\mu}_t(v) \label{mutf} \end{eqnarray} 
\end{definition}

One can remark that for $A(x,y) = \frac{1}{2}(|x|+|y|-|x-y|)$ and $v\in {\mathcal D}$, $\int\!\!\int_{\RR\times
    \RR} v(x)v(y) A(x,y)dxdy = \int_0^{+\infty}(w(x)^2 +
  (w(-\infty)-w(-x))^2)dx$ where $w(x) = \int_x^{+\infty} v(y)dy $. If
  moreover $v\in {\mathcal D_0}$,  \\$\int\!\!\int_{\RR\times
    \RR} v(x)v(y) A(x,y)dxdy=  \int_\RR w(x)^2 dx$.

A justification for this definition is the proposition, proven in
\cite{CD}, that if $u$ is a homogeneous process of finite variance,
then  
\begin{eqnarray*} 
\lefteqn{\lim_{a\rightarrow
    -\infty, b \rightarrow +\infty} \int_E  \left(\int_{\RR} \frac{i}{2}
  \left(u(x) -\frac{1}{b-a} \int_a^b u(y)dy\right) ^2 v'(x) dx\right)  \exp \, [i \int_{\RR}
u\, v\,] \,d\mu(u) =}\\ &&  \int_E  \left(\int_{\RR} i\frac{1}{2}
  u(x) ^2 v'(x) dx\right)  \exp \, [i \int_{\RR}
u\, v\,] \,d\mu(u)
\end{eqnarray*}

\section{Levy solutions}

\subsection{Evolution equation}

Now we can look for solutions that would be Levy processes with finite variance. Such  processes are  
characterized by their
exponent $\psi_t $, 
%which satisfies $\forall \, x<y, \forall \, \lambda \in
%\RR$~:
%\[
%E\{ \exp \,i [\lambda(u(y,t) - u(x,t))] \} = \exp[(y-x)\psi_t(\lambda)]
%\]
or by their Laplace exponent $\phi_t$. 

 Let us recall here a useful lemma characterizing Laplace exponents of Levy processes. 

\begin{lemma}\label{laplace}
A function $\phi:\RR^+\mapsto \RR$ is the Laplace exponent
of a homogeneous Levy process with finite variance and negative jumps if and only if $\phi$ is $C^{\infty}$ on $]0,+\infty[$, $C^2$ on $\RR^+$, $\phi(0)=0$ and $\phi''$ is completely monotonous.
\end{lemma}

Moreover, Bernstein's lemma asserts that a fonction $g$ continuous on $\RR^+$, $C^\infty$ on $]0,+\infty[$ is completely monotonous if and only if $\forall 
n \in \mathbb{N}, (-1)^n g^{(n)} \geq 0$. Injecting the definition of the exponents \ref{expo}
 in the evolution equation, one gets the following result.

%is led 
% to compute
%$  \lim_{a\rightarrow
%    -\infty, b \rightarrow +\infty} \int_E  (\int_{\RR} i\frac{1}{2}
%  (u(x) -\frac{1}{b-a} \int_a^b u(y)dy) ^2 v'(x) dx)) \exp \, [i \int_{\RR}
%u\, v\,] \,d\mu_t(u)$ in terms of $\psi_t$. 
%This was done by Carraro and Duchon,
%who 

\begin{theorem} 
Let $(\psi_t)_{t\geq 0} $ be a family of exponents of Levy
processes of finite variance, such that $\forall \lambda$ $t\mapsto
\psi_t(\lambda)$ is differentiable and $\partial_t \psi_t$ is locally
bounded. Then $\psi_t$ 
characterizes an intrinsic statistical solution of our forced Burgers equation if
and only if it satisfies the equation :
$$\forall w \in \RR, \partial_t \psi(t,w) = i\partial_w \psi(t,w)
\psi(t,w) - \frac{w^2}{2} $$
\end{theorem}

If the family of exponents has negative jumps, the equation for $\phi$
is
\begin{equation}\label{eqphi}
\partial_t \phi(t,w) =- \partial_w \phi(t,w)
\phi(t,w) + \frac{w^2}{2}
\end{equation}

\begin{proof}
Suppose that $u_t$ is a family of Levy processes of exponents 
$(t,w) \mapsto \psi(t,w)$.
If $v$ is a function in $\mathcal{D}_0$, the
definition of $\psi$ gives $\hat{{\mu}}_t(v) = \exp(\int_{\RR}
\psi(t,\int_x^{+\infty} v(y)dy)dx)$ (as can be seen by considering first
simple functions).
Hence,  thanks to local boundedness, $\partial_t \hat{{\mu}}_t(v) = \hat{{\mu}}_t(v) \int_{\RR}
\partial_t \psi(t, \int_x^{+\infty} v(y)dy)dx = \hat{{\mu}}_t(v) \int_{\RR}
\partial_t \psi(t,w(x))dx$ where we denote as before 
$w(x) = \int_x^{+\infty} v(y)dy$.
The right hand side of (\ref{stats2}) is more complicated to get. 
One needs to compute \\
$  \lim_{a\rightarrow
    -\infty, b \rightarrow +\infty} \int_E  (\int_{\RR} i\frac{1}{2}
  (u(x) -\frac{1}{b-a} \int_a^b u(y)dy) ^2 v'(x) dx)) \exp \, [i \int_{\RR}
u\, v\,] \,d\mu_t(u)$ in terms of $\psi$. This was done in \cite{CD}
where it is  shown that this limit is
%\begin{equation}
$ i (\int_{\RR} \partial_w \psi(t,
w(x))\psi(t,w(x))dx) \hat{{\mu}}_t(v)$.
%\end{equation}
The other term involves \\$\int\!\!\int_{\RR\times \RR} v(x)v(y)A(x,y) dxdy
$. We have to write it in terms of $w$ : as noticed before, the
specific form of $A$ yields $\int\!\!\int_{\RR\times \RR} v(x)v(y)A(x,y) dxdy= \int_{\RR}w^2(x)dx.$
Finally, 
we must have for all $v$ in $\mathcal{D}_0$,\\$\int_{\RR}
\partial_t \psi(t,w(x))dx  =  i \int_{\RR} \partial_w \psi(t,
w(x))\psi(t,w(x))dx - \frac{1}{2}\int_{\RR} w^2(x)dx$. This is true if and only
if\\
$\forall w \in \RR, \partial_t \psi(t,w) = i\partial_w \psi(t,w)
\psi(t,w) - \frac{w^2}{2}$.
\end{proof}

 The stationary solution of equation (\ref{eqphi}) is
$
\phi(w) = \frac{w^{\frac{3}{2}}}{\sqrt{3}}$ : this is the exponent of a Levy-stable process (which of course is not of finite variance: $\phi$ is not $C^2$ in 0). Speaking informally, this is an ``invariant measure'' of our forced Burgers equation.

%Moreover, by writing equations for $\partial_w \phi(t,w)|_{w=0}$ and $\partial^2_w \phi(t,w)|_{w=0}$ one gets $E[(u(x)-u(y))] = \frac{(x-y)}{t+A}$ and
%$E[(u(x)-u(y))^2] = (x-y)[B(t+A)^{-3} + \frac{t+A}{4}]+ \frac{(x-y)^2}{(t+A)^2}$. But one can say more.

Equation (\ref{eqphi}) can be solved using characteristics. Characteristics for
(\ref{eqphi}) are curves $(\gamma_{x_0}(s),\lambda_{x_0}(s))$ that are solutions of the system 

$$\left\{
\begin{array}{lcl}
\frac{d\gamma}{ds} &=& \lambda(s) \nonumber \\
 \frac{d\lambda}{ds} &=& \frac{\gamma(s)^2}{2} \label{char}
\end{array} \right. $$
with initial condition $\gamma_{x_0}(0) = x_0, \lambda_{x_0}(0) = \phi_0(x_0)$. Each $x_0$ corresponds to a different characteristics. To find the value
of $\phi(x,t)$, one has to find $x_0$, if it exists, such that $\gamma_{x_0}(t) = x$; if $x_0$ exists and is unique, then $\phi(x,t) = \lambda_{x_0}(t)$.  

One can notice that system (\ref{char}) describes the motion of a particle in a potential $V(x)=-\frac{x^3}{6}$, starting at $x_0$ with a velocity $\phi_0(x_0)$. Thanks to energy conservation, it is thus equivalent to
$\lambda = \gamma', \gamma^3 -3(\gamma')^2 = x_0^3 -3\phi_0(x_0)^2, \gamma(0) = x_0$. 

These remarks, and the preceding lemma characterizing Laplace exponents, are
the basic ingredients of the next result.

\begin{theorem}
Let $\phi_0$ be the exponent of a Levy process of finite variance with 
negative jumps such that $\phi_0'(0)\geq 0$. Then the equation (\ref{eqphi})
with initial condition $\phi_0$ admits a unique solution which is for all $t>0$  the exponent of a homogeneous Levy process with finite variance and negative 
jumps.
Such a Levy process is thus an intrinsic statistical solution of the
randomly forced Burgers equation.\\
If $\phi_0=0$, the solution is $\phi(x,t) = t^{-3} \sqrt{(xt^2)^3 -
  F(xt^2)^6}$ where $F$ is the inverse of 
$x\mapsto 4^{\frac{1}{3}} x^2 {\mathcal{P}}(\omega_2 +
\frac{x}{\sqrt{3}4^{\frac{1}{3}}})$. Here $\mathcal{P}$ is the Weierstrass
function and $\omega_2$ its half-period.
\end{theorem}

We will start the proof by looking at the particular case $\phi_0=0$. It is
instructive to see how things go, and it is anyway necessary to deal with it
separately. The general case will be covered afterwards.

\subsection{The particular case $\phi_0=0$}

In this particular case (corresponding to $u_0$ constant),
the system (\ref{char}) is equivalent to
$\lambda = \gamma', \gamma^3 -3(\gamma')^2 = x_0^3, \gamma(0) = x_0$. 
The solution of this equation is $\gamma(s) = 4^{\frac{1}{3}} x_0 {\mathcal{P}}(\omega_2 + \frac{\sqrt{x_0}}{\sqrt{3} 4^{\frac{1}{3}}} s)$ where ${\mathcal{P}}$ is the Weierstrass function solution of $y'^2 -4y^3 + 1=0$ (with invariants
$g_2=0$ and $g_3=1$), and $\omega_2 =
\frac{\Gamma^3(\frac{1}{3})}{4\pi}$ is the half period of
${\mathcal{P}}$, satisfying ${\mathcal{P}}'(\omega_2) = 0$. $\gamma$
is 1-1  from $[0,\frac{\sqrt{3}4^{\frac{1}{3}}}{\sqrt{x_0}} \omega_2[$
to $[x_0,+\infty[$ (see \cite{AS}); it is $C^\infty$ and strictly increasing on  $]0,\frac{\sqrt{3}4^{\frac{1}{3}}}{\sqrt{x_0}} \omega_2[$.
In order to find $\phi(x,t)$, one needs to find $x_0$ such that
$4^{\frac{1}{3}}x_0 {\mathcal{P}}(\omega_2 + \frac{\sqrt{x_0}}{\sqrt{3} 4^{\frac{1}{3}}} t)=x$. Hence $x_0 = \frac{F(xt^2)^2}{t^2}$ where $F$ is the inverse of 
$x\mapsto 4^{\frac{1}{3}} x^2 {\mathcal{P}}(\omega_2 + \frac{x}{\sqrt{3}4^{\frac{1}{3}}})$. $F$ is continuous from $[0,+\infty[$ to $[0,\sqrt{4}4^{\frac{1}{3}} \omega_2[$ and $C^{\infty}$ on $]0,+\infty[$.  Near 0, one has
 $ 4^{\frac{1}{3}} x^2 {\mathcal{P}}(\omega_2 +
 \frac{x}{\sqrt{3}4^{\frac{1}{3}}}) = x^2 + \frac{x^4}{4} + O(x^6)$,
 hence $F(x) = \sqrt{x} -\frac{x^{\frac{3}{2}}}{8} + o(x^2)$. \\
Thus the solution of (\ref{eqphi}) with initial condition $\phi_0 = 0$ is $\phi(x,t) = t^{-3} \sqrt{(xt^2)^3 - F(xt^2)^6}$; the development of $F$ near $0$   ensures that $x\mapsto \phi(x,t)$ is $C^2$ on $[0,+\infty[$.
The proof that $\phi''$ is completely monotone will be done for the general case.

\subsection{General case}

As a preliminary remark, since $\phi_0''$ is completely monotone, $\phi'_0\geq 0$  on $\RR^+$, hence $\phi_0$ is nonnegative and increasing on $\RR^+$.
Moreover, if there is an interval $I=[0,a[$ such that $\phi_0(w)=0 \forall w \in I$,  
then $\phi_0''=0$ on $I$. But $\phi''$ is decreasing (because $\phi_0^{(3)}\leq 0$) hence $\phi_0''=0$ on $\RR^+$. Therefore  $\phi_0=0$ on $\RR^+$ or $\phi_0>0$ on
$]0,+\infty[$. The particular case $\phi_0=0$ has been covered in the preceding section (except for complete monotonocity), 
hence we will assume here that $\phi_0>0$ on $\RR^{+*}$.

Everywhere in this proof $\phi(x,t)'$,$\phi(x,t)''$, $\phi(x,t)^{(n)}$ will
refer to partial derivatives with respect to $x$.

As noticed before,  the solution $\gamma_{x_0}(t)$ of the system (\ref{char}) is the trajectory of a particle in a potential $V(x)=-\frac{x^3}{6}$, starting at $x_0$ with a velocity $\phi_0(x_0)$.  Hence $\gamma_{x_0}(t)$ is a continuous
 increasing function of $x_0$. More precisely, if $x_0=0$ $\gamma_0(t)=0$, else if $x_0\neq 0$,

\begin{equation*}
\gamma_{x_0}(t)=x \Leftrightarrow t=\int_{x_0}^x \frac{dy}{\sqrt{y^3-x_0^3+3\phi_0^2(x_0)}}
\end{equation*}

The function $h:\{(x,x_0)/0< x_0\leq x\} \rightarrow \RR^+$, defined by $h(x,x_0) = \int_{x_0}^x \frac{dy}{\sqrt{y^3-x_0^3+3\phi_0^2(x_0)}}$ is $C^\infty$ ($\phi_0(x_0)>0$, therefore this
integral is well defined); $x_0\mapsto h(x,x_0)$ 
is a  diffeomorphism from $]0,x]$ into $\RR^+$ hence for all $t>0$ $\gamma_{x_0}(t) = x$ defines uniquely $x_0$ as
a $C^\infty$ function of $(x,t)$ on $]0,+\infty[^2$, satisfying $\lim_{x\rightarrow 0} x_0(x,t) = 0$. 
 
Hence the equation (\ref{eqphi}) admits as a solution for all $t$, $\phi(x,t) = 
\sqrt{x^3-x_0(x,t)^3 + 3\phi_0^2(x_0)}$, $\phi(0,t) = 0$. $x\rightarrow \phi(x,t) $ is $C^\infty$ on $]0,+\infty[$. We still have to check that $\phi''$ is 
completely monotone and that $\phi$ is $C^2$ at 0.

Let us prove that $\phi''$ is completely monotone. This proof will also work 
for $\phi_0=0$.
%First we prove that $\phi'\geq 0$. Obviously $x_0\mapsto \gamma_{x_0}(t)$ is an increasing function of $x_0$ : starting faster and further away, one arrives even further. $x_0\mapsto \lambda_{x_0}(t)$ is also an increasing function of
%$x_0$ : if $x_0<x_1$, $0\leq \lambda_{x_0}(0)<\lambda_{x_1}(0)$ and $\frac{d}{dt}(\lambda_{x_0}(t)-\lambda_{x_1}(t))\leq 0$. Hence $\phi(x,t)=\lambda(x_0(x,t),t)$ is also an increasing function of $x$. 
The keypoint is to notice that $\frac{d}{dt}(\phi(\gamma_{x_0}(t),t)) = \lambda_{x_0}(t)\phi'(\gamma_{x_0}(t),t) + \frac{\partial \phi}{\partial t}(\gamma_{x_0}(t),t) =  \phi\phi'(\gamma_{x_0}(t),t) + \frac{\partial \phi}{\partial t}(\gamma_{x_0}(t),t)$. 
Hence by deriving twice equation (\ref{eqphi}) with respect to $w$, and by
setting $w=\gamma_{x_0}(t)$ one gets
\begin{equation*}
\frac{d}{dt}(\phi''(\gamma_{x_0}(t),t)) + 3 \phi'\phi''(\gamma_{x_0}(t),t) = 1
\end{equation*}

Thus, for every $x_0$, $t\mapsto \phi''(\gamma_{x_0}(t),t))$ satisfies a differential inequality of the
form $\frac{du}{dt} +3 f u\geq 0$ (where $f = \phi'(\gamma_{x_0}(t),t)$). Since $\phi''_0 \geq 0$, we deduce $\phi''(\gamma_{x_0}(t),t)\geq 0$ for all $t>0$. Since it 
is true for all $x_0$, $\phi''(x,t)\geq 0$ for all $x>0,t>0$.

One can now proceed by induction. Suppose $(-1)^n \phi^{(n)}(x,t)\geq 0$ for all $2\leq n\leq N$. Then by deriving (\ref{eqphi}) $N+1$ times one gets
\begin{equation*}
\frac{d}{dt}(\phi^{(N+1)}(\gamma_{x_0}(t),t)) + (N+2)f \phi^{(N+1)}(\gamma_{x_0}(t),t)) =-\frac{1}{2} \sum_{k=2}^{N} \binom{N+2}{k}  \phi^{k}(\gamma_{x_0}(t),t)) \phi^{(N+2-k)}(\gamma_{x_0}(t),t))
\end{equation*}

Now the induction hypothesis guarantees that the right hand side of this
equation is of the sign of $(-1)^{N+1}$. Hence $ (-1)^{N+1} \phi^{(N+1)}(\gamma_{x_0}(t),t)$ verifies $\frac{du}{dt} +(N+2) f u\geq 0$. One concludes as before
that   $(-1)^{N+1} \phi^{(N+1)}(x,t)\geq 0$.

Hence $\phi(x,t)$ is completely monotone.

Let us now prove that $\phi$ is $C^2$ at $x=0$ : we need to know how $x_0$ behaves near $x=0$. In order to do this we
will consider the solution of
$y^3-3y'^2+C(x_0)=0, y(0)=x_0$, where $C(x_0) = 3\phi_0(x_0)^2 -x_0^3$. Its explicit expression depends on the sign of $C$. We know that $\lim_{x\rightarrow 0} x_0 =0$. Hence there are really two cases that have to be dealt with : $\phi_0'(0)>0$ and $\phi_0'(0)=0$.
\begin{itemize} 
\item {\it If $\phi_0'(0) = 0$}

$\phi_0$ is $C^2$ at 0, hence $\phi_0(x_0) = k^2 x_0^2 +o(x_0^2)$ where $k$ is a constant,  and  $C(x_0) = -x_0^3 + 3k^4 x_0^4 + o(x_0^4)$ is 
negative when $x_0$ is small enough. Hence $\gamma_{x_0}(t) = A\mathcal{P}(bt+c)$ where $A^3 = -4C(x_0)$, $b=\sqrt{\frac{A}{12}}$, $\mathcal{P}$ is as before the Weierstrass function with invariants $g_2=0,g_3=1$, and $c$ verifies $A\mathcal{P}(c) = x_0$. 
\\
From the expansion $\mathcal{P}(\omega_2 + \epsilon) = 4^{-\frac{1}{3}} + 3.4^{-\frac{2}{3}} \epsilon^2 + o(\epsilon^2)$ we deduce 
\begin{eqnarray*}
A&=&4^{\frac{1}{3}}x_0(1-k^4 x_0) + o(x_0^2)\\
b&=& \sqrt{\frac{4^{\frac{1}{3}}}{12}}\sqrt{x_0} + o(\sqrt{x_0})\\
c&=&\omega_2 + \sqrt{\frac{4^{\frac{1}{3}}}{3}}k^2 \sqrt{x_0} + o(\sqrt{x_0})\\
\gamma_{x_0}(t) &=& x_0 + (k^2t + \frac{t^2}{4})x_0^2+o(x_0^2)\\
x_0(x,t) &=&x-(k^2t+\frac{t^2}{4})x^2 + o(x^2)\\
\phi(x,t) &=& \sqrt{\frac{x^3-x_0^3+3\phi_0(x_0)^2}{3}} = (k^2 + \frac{t}{2})x^2 + o(x^2)
\end{eqnarray*}

Hence $\phi$ is $C^2$ at $x=0$.

\item {\it If $\phi'_0(0) = a > 0$}

Now $C(x_0) =3a^2 x_0^2  + k x_0^3 + o(x_0^3)$, where $k$ is a constant, is 
positive when $x_0$ is small enough. Hence $\gamma_{x_0}(t) = A\mathcal{Q}(bt+c)$ where $A^3 = 4C(x_0)$, $b=\sqrt{\frac{A}{12}}$, and $\mathcal{Q}$ is now
 the Weierstrass function with invariants $g_2=0,g_3=-1$; and $c$ must verify $A\mathcal{Q}(c) = x_0$. 
\\
$\mathcal{Q}$ is related to $\mathcal{P}$ by $\mathcal{Q}(z) = -\mathcal{P}(iz)$ for all $z$ in $\mathbb{C}$. $\mathcal{Q}$ admits a real zero $z_1$, with $\mathcal{Q}'(z_1) = 1$ and $\mathcal{Q}''(z_1) = \mathcal{Q}'''(z_1)=0$. It is not difficult to get 
the expansion $\mathcal{Q}(z_1 + \epsilon) = \epsilon + \frac{\epsilon^4}{2} + o(\epsilon^4)$, and : 
\begin{eqnarray*}
A&=&(12a^2)^{\frac{1}{3}}x_0^{\frac{2}{3}}(1+\frac{k}{9a^2} x_0 + o(x_0))\\
b&=& (\frac{ax_0}{12})^{\frac{1}{3}}(1+\frac{k}{18a^2}x_0 + o(x_0))\\
c&=&z_1 + (\frac{x_0}{12a^2})^{\frac{1}{3}} -a^{-\frac{8}{3}}(\frac{12^{-\frac{4}{3}}}{2} + k \frac{12^{-\frac{1}{3}}}{9}) x_0^{\frac{4}{3}}  + o({x_0}^{\frac{4}{3}})\\
\gamma_{x_0}(t) &=& x_0(1+at) + (\frac{k}{18a}t+ \frac{1}{24a^2}((1+at)^4-1) )x_0^2+o(x_0^2)\\
x_0(x,t) &=&\frac{x}{1+at} -  ( (\frac{k}{18a}t+ \frac{1}{24a^2}((1+at)^4-1)) \frac{x^2}{(1+at)^3} + o(x^2)\\
\phi(x,t) &=& \sqrt{\frac{x^3-x_0^3+3\phi_0(x_0)^2}{3}} = \frac{ax}{1+at} \sqrt{1+K_2 x + o(x)}\,\,\mbox{ where $K_2=\frac{1+3(1+at)^4+4k(1+\frac{2at}{3})}{12a^2(1+at)^2}$ }
\end{eqnarray*}

Hence $\phi$ is here also $C^2$ at $x=0$.

\end{itemize}
\bibliographystyle{plain}

\end{document}